\documentclass
[floatfix,superscriptaddress,secnumarabic,amssymb,amsmath,nobibnotes,aps,prd,showkeys,showpacs,nofootinbib,onecolumn,notitlepage,12pt]{revtex4}%
\usepackage{setspace}
\usepackage{color}
\usepackage{amsmath}
\usepackage{amsfonts}
\usepackage{verbatim}
\usepackage{amssymb}
\usepackage{graphicx,bm}
\usepackage{graphicx}
\usepackage{amsmath}
\usepackage{amssymb}
\usepackage{amssymb}
\usepackage{graphicx,bm}
\usepackage{graphicx}
\usepackage[caption=false]{subfig}%
\providecommand{\U}[1]{\protect\rule{.1in}{.1in}}

\newcommand{\be}{\begin{equation}}
\newcommand{\ee}{\end{equation}}

\newcommand{\mincir}{\raise
-3.truept\hbox{\rlap{\hbox{$\sim$}}\raise4.truept\hbox{$<$}\ }}
\newcommand{\magcir}{\raise
-3.truept\hbox{\rlap{\hbox{$\sim$}}\raise4.truept\hbox{$>$}\ }}

\ifx\pdfoutput\relax\let\pdfoutput=\undefined\fi
\newcount\msipdfoutput
\ifx\pdfoutput\undefined\else
\ifcase\pdfoutput\else
\ifx\paperwidth\undefined\else
\ifdim\paperheight=0pt\relax\else\pdfpageheight\paperheight\fi
\ifdim\paperwidth=0pt\relax\else\pdfpagewidth\paperwidth\fi
\fi\fi\fi
\begin{document}
\title{Semi-Classical limit and quantum corrections in noncoincidence power-law
$f(Q)$-Cosmology}
\author{Andronikos Paliathanasis}
\email{anpaliat@phys.uoa.gr}
\affiliation{Institute of Systems Science, Durban University of Technology, Durban 4000,
South Africa}
\affiliation{Departamento de Matem\'{a}ticas, Universidad Cat\'{o}lica del Norte, Avda.
Angamos 0610, Casilla 1280 Antofagasta, Chile}

\begin{abstract}
Within the framework of symmetric teleparallel $f\left(  Q\right)  $-gravity
for a connection defined in the non-coincidence gauge we derive the
Wheeler-DeWitt equation of quantum cosmology. Because the gravitational field
equation in $f\left(  Q\right)  $-gravity admits a minisuperspace description
the Wheeler-DeWitt equation is a single inhomogeneous partial differential
equations. We assume the power-law $f\left(  Q\right)  =f_{0}Q^{\mu}$ model
and with the application of linear quantum observables we calculate the
wavefunction of the universe. Finally, we investigate the effects of the
quantum correction terms in the semi-classical limit.

\end{abstract}
\maketitle

\section{Introduction}

\label{sec1}

Symmetric teleparallel general relativity
\cite{Nester:1998mp,ln1,ln2,hcov1,hcov2,revh} (STEGR) is a gravitational
theory that is equivalent to General Relativity (GR). In STEGR, the geometry
of the physical space is described by a metric tensor as in GR, but the
autoparallels are defined by a symmetric and flat connection that inherits the
symmetries of the metric tensor. This leads to different autoparallels than
those of GR, which are constructed by the Levi-Civita connection.

Assume that $\tilde{\Gamma}_{\mu\nu}^{\kappa}$ is a general connection. This
can be decomposed \cite{mt1} as $\tilde{\Gamma}_{\mu\nu}^{\kappa}=%
\begin{Bmatrix}
{\small \kappa}\ \\
{\small \mu~\nu}%
\end{Bmatrix}
+2\Gamma_{\left[  \mu\nu\right]  }^{\kappa}+\Delta_{\mu\nu}^{\kappa}$, where $%
\begin{Bmatrix}
{\small \kappa}\ \\
{\small \mu~\nu}%
\end{Bmatrix}
$ is the Levi-Civita connection, $\Gamma_{\left[  \mu\nu\right]  }^{\kappa}$
is the torsion tensor, and~$\Delta_{\mu\nu}^{\kappa}$ is the symmetric and
flat nonmetricity part~\cite{Eisenhart}. Since in STEGR, the connection is
flat, it leads to a zero-valued curvature tensor $R_{~\lambda\mu\nu}^{\kappa
}=0$ and it is symmetric, i.e., $\mathrm{T}_{\mu\nu}^{\kappa}=0$. Thus, only
the nonmetricity tensor survives, which leads to the nonmetricity scalar $Q$.
The latter scalar substitutes the Ricci scalar of the Einstein-Hilbert Action
Integral, leading to STEGR. As the Ricci scalar $R$ and the nonmetricity
scalar $Q$ differ by a boundary term~\cite{Nester:1998mp}, it follows that
STEGR is equivalent to GR.

Nowadays, GR is challenged by the analysis of recent cosmological data
\cite{rr1, Teg, Kowal, Komatsu,h01,h02}. This has led the scientific community
to introduce alternative and modified theories of gravity. Within the symmetry
and teleparallel theory, the simplest modification is the $f\left(  Q\right)
$-gravity \cite{Koivisto2,Koivisto3}, where the Lagrangian function for the
gravitational Action Integral is a nonlinear function of the nonmetricity
scalar $Q$. In the linear limit of the function $f$, the STEGR theory, with or
without the cosmological constant term, is recovered. $f\left(  Q\right)
$-gravity is the analogue in the framework of STEGR of other modifiend
$f-$theories defined by the Levi-Civita connection or in teleparalellism, see
for instance \cite{ff1,ff2,ff3,ff4,ff5,ff6,ff7,ff8} and references therein.

The dynamical degrees of freedom introduced by the nonlinear $f\left(
Q\right)  $~can be attributed to scalar fields \cite{minfq}. In the scalar
field description, $f\left(  Q\right)  $-theory is equivalent to a specific
case of the scalar nonmetricity theory, where the scalar field is nonminimally
coupled to gravity. Therefore, $f\left(  Q\right)  $-gravity has properties
similar to a Machian theory, even if the theory is not purely Machian. For
more details, we refer the reader to \cite{sc1,gg1}.

$f\left(  Q\right)  $-theory suffers from two major problems in the
cosmological perturbations of a spatially flat
Friedmann--Lema\^{\i}tre--Robertson--Walker (FLRW) background geometry. In
particular, it suffers from strong coupling and the appearance of ghosts
\cite{ppr1,ppr2}. Despite the fact that $f\left(  Q\right)  $-theory fails to
explain the global evolution of the universe, it provides a unique mechanism
leading to the construction of different toy models with interesting
applications in gravitational physics. See, for instance,
\cite{fq1,fq2,fq3,fq5,fq6,fq7,fq8,fq9,fq10,fq11,fq12,fq13,fq14,fq15} and
references therein. For other modifications of the STEGR theory, we refer the
reader to \cite{md1,md2,md3,md4}. Moreover, $f\left(  Q\right)  $-theory is
the simplest mathematical theory which can be applied to understand the
effects of the selection of the connection in modified symmetric teleparallel theories.

This investigation deals with quantum cosmology within the framwork of
$f\left(  Q\right)  $-theory and the effects of the quantum correction in the
semiclassical limit \cite{kim}. Specifically, for the FLRW geometry, we assume
a connection defined in the noncoincidence gauge where the field equations
admit a minisuperspace description. For this specific connection, the field
equations are derived from a point-like Lagrangian with three dependent
variables: the scale factor of the FLRW geometry and two scalar fields, which
attribute the degrees of freedom provided by the definition of the connection
and the nonlinear function $f\left(  Q\right)  $. For this two-scalar field
cosmological model, we write the Wheeler-DeWitt equation (WDW) of quantum
cosmology \cite{Wil,hill}. We employ the theory of symmetries of differential
equations \cite{nrev1} to define quantum observables. These are applied to
define similarity transformations and derive the wavefunction of the universe.

In the hydrodynamic description of quantum cosmology \cite{md11,qqm}, we make
use of the quantum observables to write down conservation laws for the
classical field equations and to calculate the classical solutions, because
for the power-law $f\left(  Q\right)  $-model the classical field equations
form an superintegrable Hamiltonian system. The effects of the quantum
correction terms related to the nonzero Ricci scalar term of the
minisuperspace \cite{Wil,hill}, and of the Bohmian potential term
\cite{bm1,bm2,bm4,bm5,bm6}, are discussed. It follows that the quantum
corrections do not affect the general evolution and dynamics of the classical
gravitational model. The structure of the paper is as follows.

In Section \ref{sec2} we briefly discuss STEGR and its modification, the
$f\left(  Q\right)  $-gravity.\ This gravitational theory has a scalar field
description, the theory is equivalent to a specific case of the
scalar-nonmetricity gravitational model. Furthermore, the effects of conformal
transformations in the theory are investigated, from where we can understand
the degrees of freedom introduced by the theory.

The case of an isotropic and homogeneous universe in the framework of
$f\left(  Q\right)  $-gravity is presented in Section \ref{sec3}. We review
previous results on the different families of connections in the FLRW
background, and explore how the selection of the connection affects the
gravitational model and the evolution of the physical space. Moreover, by
applying scalar fields, we introduce the minisuperspace description for the
field equations.

For the connection where the Lagrangian of the field equations is point-like,
that is, it describes the motion of classical particles in a curved space
under a conservative force term, in Section \ref{sec4}, we derive the WDW
equation, which is a single inhomogeneous partial differential equation. For
the power-law $f\left(  Q\right)  $ model in Section \ref{sec5}, we present
the quantum observables for the WDW equation. The latter are used to construct
similarity transformations aimed at reducing the WDW equation into an ordinary
differential equation and providing a closed-form expression for the
wavefunction of the universe. In Section \ref{sec6}, we employ the Madelung
representation of quantum mechanics to derive the classical and semiclassical
limits where quantum correction terms are introduced in the gravitational
field equations. For the two models, we explicitly solve the reduced
gravitational field equations using the Hamilton-Jacobi theory. It follows
that the quantum correction terms do not affect the main dynamics of the
classical solution. The two asymptotic limits of the classical cosmological
solution describe the self-similar solution related to the power-law model, as
well as a cosmological solution with two perfect fluids that do not interact
and have constant equation of state parameters. Finally, in Section
\ref{sec7}, we draw our conclusions.

\section{Symmetric teleparallel $f\left(  Q\right)  $-gravity}

\label{sec2} In this Section, we discuss the main properties and definitions
of STEGR and introduce the gravitational model under our consideration, which
is that of symmetric teleparallel $f\left(  Q\right)  $-gravity.

\subsection{STEGR}

In the framework of STEGR, the physical space is described by a four
dimensional metric tensor $g_{\mu\nu}$, and the connection $\Gamma_{\mu\nu
}^{\kappa}$ which is symmetric, i.e. $\Gamma_{\mu\nu}^{\kappa}=\Gamma_{\nu\mu
}^{\kappa}$, and flat. Hence, there exists a point transformation $x^{\mu
}\rightarrow x^{\mu^{\prime}}$, such that all the components of the connection
in the new coordinate system are zero, i.e. $\Gamma_{\mu^{\prime}\nu^{\prime}%
}^{\kappa^{\prime}}=0~$\cite{Eisenhart}.

Because of these two fundamental properties for the connection $\Gamma_{\mu
\nu}^{\kappa}$, the curvature tensor defined as
\begin{equation}
R_{\;\lambda\mu\nu}^{\kappa}\left(  \Gamma\right)  =\frac{\partial
\Gamma_{\;\lambda\nu}^{\kappa}}{\partial x^{\mu}}-\frac{\partial
\Gamma_{\;\lambda\mu}^{\kappa}}{\partial x^{\nu}}+\Gamma_{\;\lambda\nu
}^{\sigma}\Gamma_{\;\mu\sigma}^{\kappa}-\Gamma_{\;\lambda\mu}^{\sigma}%
\Gamma_{\;\mu\sigma}^{\kappa}, \label{rs.01}%
\end{equation}
and the torsion tensor which reads%
\begin{equation}
\mathrm{T}_{\mu\nu}^{\kappa}\left(  \Gamma\right)  =\frac{1}{2}\left(
\Gamma_{\;\ \mu\nu}^{\kappa}-\Gamma_{\;\ \nu\mu}^{\kappa}\right)  ,
\label{rs.02}%
\end{equation}
are always zero.

The fundamental scalar of STEGR is nonmetricity scalar \cite{Nester:1998mp}%
\begin{equation}
Q=Q_{\kappa\mu\nu}P^{\kappa\mu\nu}, \label{rs.03}%
\end{equation}
in which $Q_{\kappa\mu\nu}$ is the nonmetricity scalar~$Q_{\kappa\mu\nu
}=\nabla_{\kappa}g_{\mu\nu}$, that is,%
\begin{equation}
Q_{\kappa\mu\nu}=\frac{\partial g_{\mu\nu}}{\partial x^{\kappa}}%
-\Gamma_{\;\kappa\mu}^{\sigma}g_{\sigma\nu}-\Gamma_{\;\kappa\nu}^{\sigma
}g_{\mu\sigma}, \label{rs.04}%
\end{equation}
and $P_{\;\mu\nu}^{\kappa}$ is defined as follow%
\begin{equation}
P_{\;\mu\nu}^{\kappa}=\frac{1}{4}Q_{\;\mu\nu}^{\kappa}+\frac{1}{2}%
Q_{(\mu\phantom{\lambda}\nu)}^{\phantom{(\mu}\kappa\phantom{\nu)}}+\frac{1}%
{4}\left(  Q^{\kappa}-\tilde{Q}^{\kappa}\right)  g_{\mu\nu}-\frac{1}{4}%
\delta_{\;(\mu}^{\kappa}Q_{\nu)} \label{rs.05}%
\end{equation}
where the vector fields $Q_{\mu}$ and $\bar{Q}_{\mu}$ are given by the
expressions
\begin{equation}
Q_{\mu}=Q_{\mu\nu}^{\phantom{\mu\nu}\nu}~,~\tilde{Q}_{\mu}=Q_{\phantom{\nu}\mu
\nu}^{\nu\phantom{\mu}\phantom{\mu}}. \label{rs.06}%
\end{equation}

The gravitational Action Integral in STEGR is defined \cite{Nester:1998mp}
\begin{equation}
S=\int d^{4}x\sqrt{-g}Q.\label{rs.07}%
\end{equation}
The nonmetricity scalar $Q$ for the connection $\Gamma_{\mu\nu}^{\kappa}$ and
the Ricci scalar $\hat{R}\left(  \hat{\Gamma}\right)  $, for the Levi-Civita
connection~$\hat{\Gamma}_{\mu\nu}^{\kappa}$ of the metric tensor
\begin{equation}
\hat{\Gamma}_{\mu\nu}^{\kappa}=\frac{1}{2}g^{\kappa\lambda}\left(
g_{\mu\kappa,\nu}+g_{\lambda\nu,\mu}-g_{\mu\nu,\lambda}\right)  \label{rs.08}%
\end{equation}
are related as $Q=\hat{R}+B$, in which $B$ is a boundary term \cite{revh}~%
\begin{equation}
B=-\frac{1}{2}\mathring{\nabla}_{\lambda}P^{\lambda},\label{rs.08a}%
\end{equation}
and~$P^{\lambda}=P_{~\mu\nu}^{\lambda}\bar{g}^{\mu\nu}$. Consequently, the
variation of the Action Integral (\ref{rs.07}) leads to the field equations of
Einstein's GR which follow by the Einstein-Hilbert Action. Finally, operator
$\mathring{\nabla}_{\lambda}$ in (\ref{rs.08a}) remarks the covariant
derivative with respect to the Levi-Civita connection $\hat{\Gamma}_{\mu\nu
}^{\kappa}$ (\ref{rs.08})

\subsection{$f\left(  Q\right)  $-gravity}

The simplest modification of the STEGR Action Integral (\ref{rs.07}) is the
introduction of the cosmological constant. Nevertheless, inspired by other
modifications of GR, the introduction of nonlinear terms in the nonmetricity
scalar $Q$ in (\ref{rs.07}) leads to the introduction of dynamical degrees of
freedom which can influence the dynamics and describe the cosmic evolution.

In this general concept we assume the gravitational Action Integral to be
\cite{Koivisto2,Koivisto3}%
\begin{equation}
S_{f\left(  Q\right)  }=\int d^{4}x\sqrt{-g}f\left(  Q\right)  , \label{rs.09}%
\end{equation}
where $f\left(  Q\right)  $ is a smooth differentiable function, and when
$f^{\prime\prime}\left(  Q\right)  =0$, the limit of STEGR with or without the
cosmological constant term is recovered. With a prime $^{\prime}$ we note
total derivative with respect to the scalar $Q$, i.e. $f^{\prime}\left(
Q\right)  =\frac{df\left(  Q\right)  }{dQ}$ and $f^{\prime\prime}\left(
Q\right)  =\frac{d^{2}f\left(  Q\right)  }{dQ^{2}}$.

The modified gravitational field equations related to the Action Integral
(\ref{rs.09}) follows from the variation with respect to the metric tensor
$g_{\mu\nu}$.

The gravitational field equations are \cite{Koivisto2,Koivisto3}%
\begin{equation}
\frac{2}{\sqrt{-g}}\nabla_{\lambda}\left(  \sqrt{-g}f^{\prime}(Q)P_{\;\mu\nu
}^{\lambda}\right)  -\frac{1}{2}f(Q)g_{\mu\nu}+f^{\prime}(Q)\left(  P_{\mu
\rho\sigma}Q_{\nu}^{\;\rho\sigma}-2Q_{\rho\sigma\mu}P_{\phantom{\rho\sigma}\nu
}^{\rho\sigma}\right)  =0. \label{rs.10}%
\end{equation}
equivalently~%
\begin{equation}
f^{\prime}\left(  Q\right)  G_{\mu\nu}-\frac{1}{2}g_{\mu\nu}\left(  f\left(
Q\right)  -f^{\prime}\left(  Q\right)  Q\right)  +2f^{\prime\prime}\left(
Q\right)  P_{~~\mu\nu}^{\lambda}\nabla_{\lambda}Q=0, \label{rs.11}%
\end{equation}
where $G_{\mu\nu}$ is the equivalent of the Einstein tensor in STEGR, that is,%
\begin{equation}
G_{\mu\nu}=\left(  P_{\mu\rho\sigma}Q_{\nu}^{\;\rho\sigma}-2Q_{\rho\sigma\mu
}P_{\phantom{\rho\sigma}\nu}^{\rho\sigma}\right)  +\frac{2}{\sqrt{-g}}%
\nabla_{\lambda}\left(  \sqrt{-g}P_{\;\mu\nu}^{\lambda}\right)  .
\label{rs.12}%
\end{equation}

When $Q=Q_{0}$, expression (\ref{rs.11}) reads%
\begin{equation}
G_{\mu\nu}+\Lambda_{eff}g_{\mu\nu}=0,
\end{equation}
where $\Lambda_{eff}~$is an effective cosmological constant term defined as
$\Lambda_{eff}=\frac{1}{2}\frac{f\left(  Q_{0}\right)  -f^{\prime}\left(
Q_{0}\right)  Q_{0}}{f^{\prime}\left(  Q_{0}\right)  }$.

Therefore, for the case where $Q=Q_{0}$ and $f\left(  Q_{0}\right)
-f^{\prime}\left(  Q_{0}\right)  Q_{0}=0$, the vacuum solutions of STEGR are
recovered. On the other hand, for the case where $Q=Q_{0}$ and $f\left(
Q_{0}\right)  -f^{\prime}\left(  Q_{0}\right)  Q_{0}\neq0$, the effects of a
nonzero cosmological constant are present in STEGR.

Furthermore, varying the Action Integral (\ref{rs.09}) with respect to the
symmetric and flat connection $\Gamma_{\mu\nu}^{\kappa}$ yields the equation
of motion
\begin{equation}
\nabla_{\mu}\nabla_{\nu}\left(  \sqrt{-g}f^{\prime}%
(Q)P_{\phantom{\mu\nu}\sigma}^{\mu\nu}\right)  =0. \label{rs.13}%
\end{equation}

The connection is characterized as being defined in the \textquotedblleft
coincidence gauge\textquotedblright\ when equation (\ref{rs.13}) is
identically satisfied. On the other hand, we refer to the connection as being
defined in the \textquotedblleft noncoincidence gauge\textquotedblright%
\ \cite{Koivisto2,Koivisto3}.

The transformation rule for the metric tensor follows that of a tensor field.
However, the connection is not a tensor and thus has a different
transformation rule. Specifically, the connection is coordinate dependent,
though the tensors defined by the connection, such as the curvature tensor,
the torsion tensor, and the nonmetricity tensor, are coordinate-independent.

When we consider a specific line element for the metric tensor, this
corresponds to the definition of a proper coordinate system. Thus, the
components of the connection may not be identically zero in these coordinates,
and equation (\ref{rs.13}) is not trivially satisfied. This means that
dynamical degrees of freedom are introduced due to the connection defined in
the non-coincidence gauge. These dynamical degrees of freedom associated with
the non-coincidence gauge are of geometric origin. As we shall see in the
following section, they can be attributed to scalar fields.

\subsection{Equivalency with scalar field theory}

We introduce the Lagrange multiplier $\lambda_{m}$, such that the
gravitational Action Integral (\ref{rs.09}) reads%
\begin{equation}
S_{f\left(  Q\right)  }=\int d^{4}x\sqrt{-g}\left(  f\left(  Q\right)
+\lambda_{m}\left(  Q-\hat{Q}\right)  \right)  , \label{rs.14}%
\end{equation}
where $\hat{Q}=\hat{Q}\left(  x^{\kappa}\right)  $ and is the functional
expression for the nonmetricity scalar $Q$.

Variation with respect to the scalar $Q$, leads to the equation of
motion~$\frac{\delta S_{f\left(  Q\right)  }}{\delta Q}=0,~$that is,
$\lambda_{m}=-f^{\prime}\left(  Q\right)  $.

By replacing in expression (\ref{rs.14}) it follows%
\begin{equation}
S_{f\left(  Q\right)  }=\int d^{4}x\sqrt{-g}\left(  f^{\prime}\left(
Q\right)  \hat{Q}+\left(  f\left(  Q\right)  -Qf^{\prime}\left(  Q\right)
\right)  \right)  . \label{rs.15}%
\end{equation}
We introduce the scalar field $\phi=f^{\prime}\left(  Q\right)  $, and the
potential function $V\left(  \phi\right)  =\left(  Qf^{\prime}\left(
Q\right)  -f\left(  Q\right)  \right)  $, and the latter Action Integral is
expressed in the following simpler form%
\begin{equation}
S_{f\left(  Q\right)  }=\int d^{4}x\sqrt{-g}\left(  \phi\hat{Q}-V\left(
\phi\right)  \right)  . \label{rs.16}%
\end{equation}

This is analogous to O'Hanlon gravity \cite{ohn1} in the framework of
symmetric teleparallel theory. Hence, we can say that the nonmetricity
$f\left(  Q\right)  $-gravity has properties similar to those of a Machian
theory~\cite{revmach}, although the theory is not purely Machian
\cite{sc1,gg1}.

The Action Integral (\ref{rs.16}) is a particular case of the more general
theory with a scalar field minimally coupled to gravity, that is, of the
scalar-nonmetricity theory defined as
\begin{equation}
S_{ST\phi}=\int d^{4}x\sqrt{-g}\left(  \frac{F\left(  \phi\right)  }{2}%
Q-\frac{\omega\left(  \phi\right)  }{2}g^{\mu\nu}\phi_{,\mu}\phi_{,\nu
}-V\left(  \phi\right)  \right)  , \label{rs.17}%
\end{equation}
where the gravitational field equations are%
\begin{equation}
F\left(  \phi\right)  G_{\mu\nu}+2F_{,\phi}\phi_{,\lambda}P_{~~\mu\nu
}^{\lambda}+g_{\mu\nu}V\left(  \phi\right)  +\frac{\omega\left(  \phi\right)
}{2}\left(  g_{\mu\nu}g^{\lambda\kappa}\phi_{,\lambda}\phi_{,\kappa}%
-\phi_{,\mu}\phi_{,\nu}\right)  =0. \label{rs.18}%
\end{equation}

The equation of motion for the scalar field reads%
\begin{equation}
\frac{\omega\left(  \phi\right)  }{\sqrt{-g}}g^{\mu\nu}\partial_{\mu}\left(
\sqrt{-g}\partial_{\nu}\phi\right)  +\frac{\omega_{,\phi}}{2}g^{\lambda\kappa
}\phi_{,\lambda}\phi_{,\kappa}+\frac{1}{2}F_{,\phi}Q-V_{,\phi}=0,
\label{rs.19}%
\end{equation}
while the equation of motion for the connection is
\begin{equation}
\nabla_{\mu}\nabla_{\nu}\left(  \sqrt{-g}F\left(  \phi\right)
P_{\phantom{\mu\nu}\sigma}^{\mu\nu}\right)  =0. \label{rs.20}%
\end{equation}

We remark that for a linear function $F\left(  \phi\right)  =2\phi$, and
$\omega\left(  \phi\right)  =0$, the $f\left(  Q\right)  $-gravity is
recovered, and the field equations (\ref{rs.11}) become%
\begin{equation}
\phi G_{\mu\nu}+2\phi_{,\lambda}P_{~~\mu\nu}^{\lambda}+g_{\mu\nu}V\left(
\phi\right)  =0. \label{rs.21}%
\end{equation}

Moreover, equation (\ref{rs.19}) is simultaneously satisfied by the definition
of the scalar field potential $V\left(  \phi\right)  $, while the equation of
motion for the connection is simplified to
\begin{equation}
\nabla_{\mu}\nabla_{\nu}\left(  \sqrt{-g}\phi P_{\phantom{\mu\nu}\sigma}%
^{\mu\nu}\right)  =0. \label{rs.22}%
\end{equation}

\subsection{Conformal transformation}

\label{sec2a}

In the previous lines, we learned that $f\left(  Q\right)  $ is a partially
Machian theory, where, in the scalar field description, the theory is defined
in the so-called Jordan frame. In the following lines, we discuss the effects
of conformal transformations in $f\left(  Q\right)  $-gravity and introduce
the equivalent theory in the Einstein frame.

Let the two conformally related four-dimensional metrics be $\bar{g}_{\mu\nu}$
and $g_{\mu\nu}$, defined as
\begin{equation}
\bar{g}_{\mu\nu}=e^{2\Omega\left(  x^{\kappa}\right)  }g_{\mu\nu}~~,~\bar
{g}^{\mu\nu}=e^{-2\Omega\left(  x^{\kappa}\right)  }g^{\mu\nu}, \label{rs.23}%
\end{equation}
where $\Omega\left(  x^{\kappa}\right)  $ is a smooth differentiable function.
$\Omega\left(  x^{\kappa}\right)  $ is known as the conformal factor and
defines the transformations.

The geometric quantities, which define the $f\left(  Q\right)  $-gravity, for
the conformally related metrics are related as \cite{gg1}
\begin{equation}
\bar{Q}_{\lambda\mu\nu}=e^{2\Omega}Q_{\lambda\mu\nu}+2\Omega_{,\lambda}\bar
{g}_{\mu\nu}. \label{rs.24}%
\end{equation}
and%
\begin{equation}
\bar{P}^{\lambda}=\bar{P}_{~\mu\nu}^{\lambda}\bar{g}^{\mu\nu}=e^{-2\Omega
}P^{\lambda}+3\Omega^{,\lambda}. \label{rs.25}%
\end{equation}

Therefore, the nonmetricity scalars $Q$ and $\bar{Q}$ are
\begin{equation}
\bar{Q}=\bar{Q}_{\lambda\mu\nu}\bar{P}^{\lambda\mu\nu}=e^{-2\Omega}Q+\left(
2\Omega_{,\lambda}P^{\lambda}+6\Omega_{\lambda}\Omega^{,\lambda}\right)  .
\label{rs.26}%
\end{equation}

Assume now the Action Integral (\ref{rs.16}) for the $f\left(  \bar{Q}\right)
$-theory~in the framework of the space with metric $\bar{g}_{\mu\nu}$. Then,
the equivalent theory for the metric tensor $g_{\mu\nu}$ is given by the
Action Integral \cite{stn3}%
\begin{equation}
\bar{S}_{f\left(  \bar{Q}\right)  }=\int d^{n}x\sqrt{-g}\left(  Q-\frac{1}%
{2}B\ln\phi+\frac{3}{2\phi}g^{\mu\nu}\phi_{,\mu}\phi_{,\nu}-\frac{V\left(
\varphi\right)  }{\phi^{2}}\right)  ,\label{rs.27}%
\end{equation}
where we have assumed that $\Omega=-\frac{1}{2}\ln\phi$, and scalar $B$ is the
boundary term which relates the Ricci scalar $\hat{R}\left(  \hat{\Gamma
}\right)  $ for the Levi-Civita connection with the nonmetricity scalar, given
by expression (\ref{rs.08a}). Indeed, the boundary term is defined as
\cite{stn3}~$B=-\frac{1}{2}\mathring{\nabla}_{\lambda}P,$ equivalently,
\cite{stn3}
\begin{equation}
B=\left(  \tilde{Q}^{\kappa}-Q^{\kappa}\right)  .
\end{equation}

We observe that the boundary term $B$ plays an important role in the conformal
equivalent description of the theory, and it is another scalar field. The
definition of the boundary term is directly related to the definition of the
connection $\Gamma$ and the equation of motion (\ref{rs.13}).

At this point, it is important to mention that the conformal equivalency of
the gravitational theory in the two frames relates only to the trajectory
solutions for the field equations. In general, it is not an equivalence of
physical properties. Nevertheless, for the scalar-nonmetricity theory, it has
been found that conformal transformations preserve the main eras of the
cosmological history and evolution.

\section{Isotropic and homogeneous cosmology}

\label{sec3}

In cosmological scales the physical space is assumed that it is homogeneous
and isotropic described by the FLRW line element. The latter in spherical
coordinates is expressed as follows
\begin{equation}
ds^{2}=-N(t)^{2}dt^{2}+a(t)^{2}\left[  \frac{dr^{2}}{1-kr^{2}}+r^{2}\left(
d\theta^{2}+\sin^{2}\theta d\varphi^{2}\right)  \right]  , \label{rs.28}%
\end{equation}
in which $k$ is the spatial curvature of the three-dimensional space;
$N\left(  t\right)  $ is the lapse function, $a\left(  t\right)  $ is the
scale factor; and $H=\frac{1}{N}\frac{\dot{a}}{a}$,~$\dot{a}=\frac{da}{dt}$,
is the Hubble function. We assume the comoving observer $u^{\mu}=\frac{1}%
{N}\delta_{t}^{\mu}$, thus, the Hubble function is related to the expansion
rate $\theta=u_{;\mu}^{\mu}$ by the expression $H=\frac{1}{3}\theta$.

\subsection{Symmetries}

In the coordinate system $\left(  t,r,\theta,\varphi\right)  $, the six
isometries of the line element (\ref{rs.28}) are as follows
\[
K^{1}=\sin\varphi\partial_{\theta}+\frac{\cos\varphi}{\tan\theta}%
\partial_{\varphi},\quad K^{2}=-\cos\varphi\partial_{\theta}+\frac{\sin
\varphi}{\tan\theta}\partial_{\varphi},\quad K^{3}=\partial_{\varphi},
\]%
\[
K^{4}=\sqrt{1-kr^{2}}\left(  \sin\theta\cos\varphi\partial_{r}+\frac{1}%
{r}\left(  \cos\theta\cos\varphi\partial_{\theta}-\frac{\sin\varphi}%
{\sin\theta}\right)  \partial_{\varphi}\right)  ,
\]%
\begin{align*}
K^{5}  &  =\sqrt{1-kr^{2}}\left(  \sin\theta\sin\varphi\partial_{r}+\frac
{1}{r}\left(  \cos\theta\sin\varphi\partial_{\theta}+\frac{\cos\varphi}%
{\sin\theta}\right)  \partial_{\varphi}\right)  ,\\
K^{6}  &  =\sqrt{1-kr^{2}}\left(  \cos\theta\partial_{r}-\frac{\sin\theta}%
{r}\partial_{\varphi}\right)  .
\end{align*}
For $k=0$, the latter symmetry vectors form the $E^{3}\otimes SO\left(
2\right)  $ Lie algebra, otherwise for $k\neq0$, the six symmetry vectors are
the elements of the $SO\left(  4\right)  $ Lie algebra.

The requirement the connection to inherit the symmetry of the background
geometry provides the following set of constraints%
\begin{equation}
\mathcal{L}_{K^{I}}\Gamma_{\mu\nu}^{\kappa}=0, \label{rs.29}%
\end{equation}
where $\mathcal{L}_{K^{I}}$ is the Lie derivative with respect the vector
field $K^{I}$ and $I=1,2,3,4,5,6$.

The definition of the Lie derivative depends on the transformation rule.
Because the connection has a different transformation rule from that of tensor
fields, the Lie derivative differs accordingly. Specifically, in terms of
coordinates, the Lie derivative of the connection reads%

\begin{equation}
L_{K^{I}}\Gamma_{\mu\nu}^{\kappa}=K_{,~\mu\nu}^{I~\kappa}+\Gamma_{\mu\nu
,r}^{\kappa}K^{I}{}^{~r}-K^{I}{}_{,r}^{~\kappa}\Gamma_{\mu\nu}^{r}%
+K^{I}~_{,\mu}^{s}\Gamma_{s\nu}^{\kappa}+K^{I}~_{,\nu}^{s}\Gamma_{\mu
s}^{\kappa}. \label{rs.30}%
\end{equation}
Because the connection is symmetric, expression (\ref{rs.30}) is simplified
as
\begin{equation}
L_{K^{I}}\Gamma_{\mu\nu}^{\kappa}=\nabla_{\nu}\nabla_{\mu}\left(
K^{I}~^{\kappa}\right)  -R_{\mu\nu\lambda}^{\kappa}K^{I}~^{\lambda}.
\label{rs.31}%
\end{equation}
However, in symmetric teleparallel theory, the connection is flat, that is,
\begin{equation}
R_{\mu\nu\lambda}^{\kappa}\left(  \Gamma\right)  =0. \label{rs.32}%
\end{equation}

onsequently, the requirement for the connection to inherit the symmetries of
the background space is equivalent to the set of differential equations%

\begin{equation}
\nabla_{\nu}\nabla_{\mu}\left(  K^{I}~^{\kappa}\right)  =0. \label{rs.33}%
\end{equation}

\subsection{Symmetric and flat connection}

For the FLRW line element (\ref{rs.28}) and the requirements (\ref{rs.32}),
(\ref{rs.33}) it follows that there are found different families of symmetric
connections $\Gamma_{\mu\nu}^{\kappa}$. One family of connection is defined
for $k\neq0$ and three families of connections are defined for $k=0~$%
\cite{Heis2,Zhao}.

The common nonzero components for the four families of connections are
\begin{align*}
\Gamma_{~tr}^{r}  &  =\Gamma_{~rt}^{r}=\Gamma_{~t\theta}^{\theta}%
=\Gamma_{~\theta t}^{\theta}=\Gamma_{~t\varphi}^{\varphi}=\Gamma_{~\varphi
t}^{\varphi}=-\frac{k}{\gamma\left(  t\right)  },\\
\Gamma_{~rr}^{r}  &  =\frac{kr}{1-kr^{2}}~,~\Gamma_{\theta\theta}%
^{r}=-r\left(  1-kr^{2}\right)  ~,~\Gamma_{~\varphi\varphi}^{r}=-r\sin
^{2}\theta\left(  1-\kappa r^{2}\right)  ,\\
\Gamma_{\;r\theta}^{\theta}  &  =\Gamma_{\;\theta r}^{\theta}=\Gamma
_{\;r\varphi}^{\varphi}=\Gamma_{\;\varphi r}^{\varphi}=\frac{1}{r}%
~,~\Gamma_{\varphi\varphi}^{\theta}=-\sin\theta\cos\theta~,~\Gamma
_{\theta\varphi}^{\varphi}=\Gamma_{\varphi\theta}^{\varphi}=\cot\theta.
\end{align*}

For $k\neq0$, the additional nonzero components for the family $\Gamma^{k}%
~$are%
\[
\Gamma_{\;tt}^{t}=-\frac{k+\dot{\gamma}(t)}{\gamma(t)},\quad\Gamma_{\;rr}%
^{t}=\frac{\gamma(t)}{1-kr^{2}}\quad\Gamma_{\;\theta\theta}^{t}=\gamma
(t)r^{2},\quad\Gamma_{\;\varphi\varphi}^{t}=\gamma(t)r^{2}\sin^{2}(\theta),
\]
by comparing the latter connections with the notation presented in
\cite{Heis2}, it follows that $\gamma\left(  t\right)  =C_{2}\left(  t\right)
$.

Moreover, for $k=0\,$, connection $\Gamma^{A}$ has the additional nonzero
components
\[
\Gamma_{\;tt}^{t}=\gamma(t),
\]
where $\Gamma^{A}$ is connection $\Gamma_{Q}^{\left(  III\right)  }$ in
\cite{Heis2} and $\gamma\left(  t\right)  =C_{1}\left(  t\right)  .$

Connection $\Gamma^{B}$ is defined for $k=0$, with the nonzero components
\[
\Gamma_{\;tt}^{t}=\frac{\dot{\gamma}(t)}{\gamma(t)}+\gamma(t),\quad
\Gamma_{\;tr}^{r}=\Gamma_{\;rt}^{r}=\Gamma_{\;t\theta}^{\theta}=\Gamma
_{\;\theta t}^{\theta}=\Gamma_{\;t\varphi}^{\varphi}=\Gamma_{\;\varphi
t}^{\varphi}=\gamma(t),
\]
and it is connection $\Gamma_{Q}^{\left(  I\right)  }$ of \cite{Heis2} with
$\gamma\left(  t\right)  =C_{3}\left(  t\right)  $

Finally, for $k=0$, family $\Gamma^{C}$ has the nonzero components%
\[
\Gamma_{\;tt}^{t}=-\frac{\dot{\gamma}(t)}{\gamma(t)},\quad\Gamma_{\;rr}%
^{t}=\gamma(t),\quad\Gamma_{\;\theta\theta}^{t}=\gamma(t)r^{2},\quad
\Gamma_{\;\varphi\varphi}^{t}=\gamma(t)r^{2}\sin^{2}\theta,
\]
which is compared with connection $\Gamma_{Q}^{\left(  II\right)  }$ of
\cite{Heis2} and $\gamma\left(  t\right)  =C_{2}\left(  t\right)  $.

The connection $\Gamma^{A}$ is the unique connection defined in the
coincidence gauge, and the function $\gamma\left(  t\right)  $ plays no role
in the gravitational field equations. Nevertheless, the remaining three
families of connections, $\Gamma^{k}$, $\Gamma^{B}$, and $\Gamma^{C}$, are
defined in the noncoincidence gauge. In the framework of $f\left(  Q\right)
$-gravity, the function $\gamma\left(  t\right)  $ is constrained by the
equation of motion (\ref{rs.13}).

\subsection{Field equations in $f\left(  Q\right)  $-gravity}

In the following lines, we present the field equations for $f\left(  Q\right)
$-gravity in an FLRW background geometry. The selection of the connection is
crucial in the theory; thus, for each family of connections, a different set
of field equations results.

For nonzero spatial curvature and connection $\Gamma^{k}$ we calculate the
nonmetricity scalar
\begin{equation}
Q\left(  \Gamma^{k}\right)  =-\frac{6\dot{a}^{2}}{N^{2}a^{2}}+\frac{3\gamma
}{a^{2}}\left(  \frac{\dot{a}}{a}+\frac{\dot{N}}{N}\right)  +\frac
{3\dot{\gamma}}{a^{2}}+k\left(  \frac{6}{a^{2}}+\frac{3}{\gamma N^{2}}\left(
\frac{\dot{N}}{N}+\frac{\dot{\gamma}}{\gamma}-\frac{3\dot{a}}{a}\right)
\right)  .
\end{equation}
The gravitational field equations in $f\left(  Q\right)  $-gravity are%
\begin{equation}
0=3f^{\prime}(Q)H^{2}+\frac{1}{2}\left(  f(Q)-Qf^{\prime}(Q)\right)
-\frac{3\gamma\dot{Q}f^{\prime\prime}(Q)}{2a^{2}}+3k\left(  \frac{f^{\prime
}(Q)}{a^{2}}-\frac{\dot{Q}f^{\prime\prime}(Q)}{2\gamma N^{2}}\right)  ,
\end{equation}%
\begin{align}
0  &  =-\frac{2}{N}\left(  f^{\prime}(QH\right)  ^{\cdot}-3H^{2}f^{\prime
}(Q)-\frac{1}{2}\left(  f(Q)-Qf^{\prime}(Q)\right) \nonumber\\
&  +\frac{\gamma\dot{Q}f^{\prime\prime}(Q)}{2a^{2}}-k\left(  \frac{f^{\prime
}(Q)}{a^{2}}+\frac{3\dot{Q}f^{\prime\prime}(Q)}{2\gamma N^{2}}\right)  ,
\end{align}
and the equation of motion for the connection
\begin{equation}
0=\left(  f^{\prime\prime}\dot{Q}\right)  ^{\cdot}\left(  1+k\frac{a^{2}%
}{N^{2}\gamma^{2}}\right)  +f^{\prime\prime}\dot{Q}\left(  \left(
1+3k\frac{a^{2}}{N^{2}\gamma^{2}}\right)  NH+\left(  1-k\frac{a^{2}}%
{N^{2}\gamma^{2}}\right)  \frac{\dot{N}}{N}+2\frac{\dot{\gamma}}{\gamma
}\right)  .
\end{equation}

We remark that in the previous field equations, if we set $k=0$, we recover
the gravitational field equations for the connection $\Gamma^{C}$.

Furthermore, for connection $\Gamma^{A}$ defined in the coincidence gauge, the
nonmetricity scalar is
\begin{equation}
Q\left(  \Gamma^{A}\right)  =-6H^{2},
\end{equation}
and the gravitational field equations are
\begin{subequations}
\label{feq11}%
\begin{align}
0  &  =3H^{2}f^{\prime}(Q)+\frac{1}{2}\left(  f(Q)-Qf^{\prime}(Q)\right)  ,\\
0  &  =-\frac{2}{N}\left(  f^{\prime}\left(  Q\right)  H\right)  ^{\cdot
}-3H^{2}f^{\prime}(Q)-\frac{1}{2}\left(  f(Q)-Qf^{\prime}(Q)\right)  .
\end{align}

Finally, for the family of connection $\Gamma^{B}$ the nonmetricity scalar
reads%
\end{subequations}
\begin{equation}
Q=-6H^{2}+\frac{3\gamma}{N}\left(  3H-\frac{\dot{N}}{N^{2}}\right)
+\frac{3\dot{\gamma}}{N^{2}},
\end{equation}
while the gravitational field equations are
\begin{subequations}
\label{feq12}%
\begin{align}
0 &  =3H^{2}f^{\prime}(Q)+\frac{1}{2}\left(  f(Q)-Qf^{\prime}(Q)\right)
+\frac{3\gamma\dot{Q}f^{\prime\prime}(Q)}{2N^{2}}\label{rs.41}\\
0= &  -\frac{2}{N}\left(  f^{\prime}\left(  Q\right)  H\right)  ^{\cdot
}-3H^{2}f^{\prime}(Q)-\frac{1}{2}\left(  f(Q)-Qf^{\prime}(Q)\right)
+\frac{3\gamma\dot{Q}f^{\prime\prime}(Q)}{2N^{2}}.\label{rs.43}%
\end{align}
Because $\Gamma^{B}$ is defined in the noncoincidence gauge, the equation of
motion for the connection is as below%
\end{subequations}
\begin{equation}
0=\left(  f^{\prime\prime}\dot{Q}\right)  ^{\cdot}+N\dot{Q}\left(
3H-\frac{\dot{N}}{N^{2}}\right)  f^{\prime\prime}(Q).\label{rs.44}%
\end{equation}

\subsection{Minisuperspace description}

A novel property of these cosmological models is that they admit a
\textquotedblleft minisuperspace\textquotedblright\ description. In
particular, for each connection, there exists a Lagrangian function whose
variation leads to the corresponding field equations.

For connections $\Gamma^{k}$ (and $\Gamma^{C}$ for $k=0$) the corresponding
field equations follows from the variation of the (non-canonical) Lagrangian
function%
\begin{equation}
L\left(  \Gamma^{k}\right)  =-\frac{3}{N}a\phi\dot{a}^{2}+3kNa\phi-\frac{N}%
{2}V\left(  \phi\right)  +\frac{3k}{2N}a^{3}\dot{\Psi}\dot{\phi}-\frac{3}%
{2}\frac{aN}{\dot{\Psi}}\dot{\phi}, \label{rs.45}%
\end{equation}
in which $\phi=f^{\prime}\left(  Q\right)  $,~$V\left(  \phi\right)  =\left(
Qf^{\prime}\left(  Q\right)  -f\left(  Q\right)  \right)  $ and $\dot{\Psi
}=\frac{1}{\gamma}$. \ We observe that the latter Lagrangian has kinetic
components in the denominator.

Similarly, for the connection $\Gamma^{A}$ the Lagrangian function is
\begin{equation}
L\left(  \Gamma^{A}\right)  =-\frac{6}{N}a\phi\dot{a}^{2}-Na^{3}V\left(
\phi\right)  \text{.} \label{rs.46}%
\end{equation}

Finally for connection $\Gamma^{B}$ the corresponding Lagrangian of the field
equations is%
\begin{equation}
L\left(  \Gamma^{B}\right)  =-\frac{3}{N}\phi a\dot{a}^{2}-\frac{3}{2N}%
a^{3}\dot{\phi}\dot{\psi}-\frac{N}{2}a^{3}V\left(  \phi\right)  ,
\label{rs.47}%
\end{equation}
where now the scalar field $\psi$ is related to the connection as $\dot{\psi
}=\gamma$. Lagrangian function (\ref{rs.47}) is of the form of a point-like
dynamical system. In particular, it describes a constraint dynamical systems
where the scale factor and two scalar fields play the role of the particles
with interaction $U_{eff}=a^{3}V\left(  \phi\right)  $.

The existence of the lapse function in the Lagrangians is necessary in order
to reconstruct the constraint equation in each case.

Between the above Lagrangian functions, only these who correspond on
connections $\Gamma^{A}$ and $\Gamma^{B}$ describe dynamical system where
expressed in the form $L=\frac{1}{N}K_{E}-NU_{eff}$, where $K$ is the kinetic
energy, $K_{E}=\frac{1}{2}\mathcal{G}_{AB}\dot{q}^{A}\dot{q}^{B}$ and
$U_{eff}$ is the effective potential. Thus, for these two dynamical systems we
can write the Hamiltonian function in the form%
\begin{equation}
H\equiv N\left(  K_{E}+U_{eff}\right)  =0. \label{rs.48}%
\end{equation}

We make use of this property in order to continue with the quantization of the
field equations. The quantization of the field equations which correspond to
the connection $\Gamma^{A}$ has been studied in detailed in . Because of the
nature of the Lagrangian, Dirac's method for the quantization of constraint
dynamical systems was applied in \cite{qdima1}. Due to the existence of second
class constraints the quantization process was different from the usual WDW
formalism, where only first class constraints exist. See also the more recent
study \cite{qdima2}.

The field equations for the connection $\Gamma_{B}$ described by the
point-like Lagrangian function (\ref{rs.47}) possess only first class
constraints, which means that the quantization process is similar to that of
the WDW formalism and different from that of the corresponding system for the
connection $\Gamma_{A}$.

\section{The Wheeler-DeWitt equation}

\label{sec4}

In GR the Wheeler-DeWitt equation follows from the quantization of the
constraint equation in the ADM formalism (for discussion we refer the reader
to \cite{Wil,hill,alk}). The WDW equation is a hyperbolic functional
differential equation on superspace and represents a family of differential
equations at different points. However, when there exist a minisuperspace
description, the WDW reduces to a single differential equation.

In the minisuperspace description the Hamiltonian constraint is expressed as%
\begin{equation}
H=N\left[  \frac{1}{2}\mathcal{G}^{AB}P_{A}P_{B}+\mathcal{U}(\mathbf{{q}%
)}\right]  =N\mathcal{H}\equiv0,
\end{equation}
in which $\mathcal{G}^{AB}$ is the minisuperspace.

The classical field equations are invariant under conformal transformations.
This property should also hold for the WDW equation of quantum cosmology. The
quantization $P_{A}=i\hbar\partial_{A}$ leads to a differential equation of
Klein-Gordon type, which is generally not conformally invariant. Therefore,
the Klein-Gordon equation is replaced by the Yamabe equation to ensure
conformal invariance. This is achieved by modifying the potential with the
term $\frac{n-2}{8(n-1)}\mathcal{R}$, where $\mathcal{R}$ is the Ricci scalar
of the minisuperspace $\mathcal{G}^{AB}$ with dimension $n$.

The WDW equation reads%
\begin{equation}
\mathcal{H}\Psi=\left[  \hbar^{2}\left(  \frac{1}{2}\Delta-\frac{n-2}%
{8(n-1)}\mathcal{R}\right)  -\mathcal{U}(\mathbf{q})\right]  \Phi
(\mathbf{q})=0, \label{Int5}%
\end{equation}
in which $\Delta=\frac{1}{\sqrt{-\mathcal{G}}}\partial_{A}\left(
\sqrt{-\mathcal{G}}\mathcal{G}^{AB}\partial_{B}\right)  $ is the Laplace
operator with respect to the minisuperspace $\mathcal{G}^{AB}$.

As we shall see in the following a nonzero Ricci scalar $\mathcal{R}$
introduces to the Hamiltonian function of the semiclassical system a potential
term which correspond to quantum corrections.

\subsection{Connection $\Gamma^{B}$}

From Lagrangian function (\ref{rs.47}) we calculate the three-dimensional
minisuperspace%
\begin{equation}
\mathcal{G}_{AB}=%
\begin{pmatrix}
6\phi a & 0 & 0\\
0 & 0 & \frac{3}{2}a^{3}\\
0 & \frac{3}{2}a^{3} & 0
\end{pmatrix}
,~\mathcal{G}^{AB}=%
\begin{pmatrix}
\frac{1}{6\phi a} & 0 & 0\\
0 & 0 & \frac{2}{3a^{3}}\\
0 & \frac{2}{3a^{3}} & 0
\end{pmatrix}
.
\end{equation}
Thus, the corresponding Hamiltonian function is expressed as
\begin{equation}
\mathcal{H}=\frac{1}{12\phi a}P_{a}^{2}+\frac{2}{3a^{3}}P_{\phi}P_{\psi}%
-a^{3}V\left(  \phi\right)  , \label{rs.50}%
\end{equation}
with
\begin{equation}
P_{a}=6\phi a\dot{a}~,~P_{\phi}=\frac{3}{2}a^{3}\dot{\psi}~\text{and }P_{\psi
}=\frac{3}{2}a^{3}\dot{\phi}\text{.} \label{rs.51}%
\end{equation}

Moreover, the Ricci scalar for the minisuperspace is derived%
\begin{equation}
\mathcal{R}=-\frac{3}{4a^{3}\phi}, \label{rs.52}%
\end{equation}
Therefore, the WDW equation (\ref{Int5}) for the Hamiltonian function
(\ref{rs.50}), $\mathcal{H}\Phi=0$, is expressed as%
\begin{equation}
\frac{1}{6a\phi}\Phi_{,aa}+\frac{4}{3a^{3}}\Phi_{,\phi\psi}+\frac{5}%
{12a^{2}\phi}\Phi_{,a}+\frac{1}{3\phi a^{3}}\Phi_{,\psi}+\left(  \frac
{3}{32a^{3}\phi}-\frac{2}{\hbar^{2}}a^{3}V\left(  \phi\right)  \right)
\Phi=0. \label{rs.53}%
\end{equation}

\subsection{Power-law theory}

We assume that the function $f\left(  Q \right)  $ is a power-law, i.e.,
$f\left(  Q \right)  = f_{0} Q^{\mu}$. The power-law function provides a
cosmological history that depends on the selection of the connection
\cite{anfqd}. However, for all connections, the power-law $f\left(  Q \right)
$ function is associated with the existence of self-similar cosmological
solutions \cite{ndself}. For the cosmological model defined by the connection
$\Gamma^{B}$, the self-similar scaling solution is always unstable, while the
unique attractor describes the accelerated de Sitter universe. This physical
property remains the same in the presence of matter \cite{anfqm}.

In the scalar field description of the $f\left(  Q\right)  $ theory, the
power-law model $f\left(  Q\right)  =f_{0}Q^{\mu}$ leads to the power-law
potential function $V\left(  \phi\right)  =V_{0}\phi^{\kappa}$, where
constants $V_{0}$ and $\kappa$ are related with $f_{0}$,~$\mu$ as
\begin{equation}
V_{0}=\left(  \mu-1\right)  f_{0}^{-\frac{1}{\mu-1}}\mu^{-\frac{\mu}{\mu-1}%
}~,~\kappa=\frac{\mu}{\mu-1}~.
\end{equation}
Recall that for values of $\mu$ close to~$1$, the gravitational model
describes small deviations from General Relativity. We refer the reader to the
interesting discussion on the power-law $f\left(  R\right)  $ theory presented
in \cite{tcl}.

Therefore, the WDW equation (\ref{rs.53}) for the power-law model is
simplified as%
\begin{equation}
\frac{1}{6a\phi}\Phi_{,aa}+\frac{4}{3a^{3}}\Phi_{,\phi\psi}+\frac{5}%
{12a^{2}\phi}\Phi_{,a}+\frac{1}{3a^{3}\phi}\Phi_{,\psi}+\left(  \frac
{3}{32a^{3}\phi}-\frac{V_{0}}{\hbar^{2}}a^{3}\phi^{\kappa}\right)  \Phi=0.
\label{rs.54}%
\end{equation}
The solution of the WDW equation (\ref{rs.54}) provides the wavefunction
$\Phi\left(  a,\phi,\psi\right)  $ for this specific cosmological model.

The WDW equation is a linear inhomogeneous partial differential equation. In
the following lines we employ the method of similarity transformations, also
known as Lie symmetry analysis, such that to determine expressions for the
wavefunction $\Phi$ which satisfy the partial differential equation
(\ref{rs.54}).

\section{Similarity transformations}

\label{sec5}

Lie symmetry analysis is a robust method for the study of nonlinear
differential equations. It provides a systematic way to investigate the
algebraic properties and compute solutions for nonlinear and inhomogeneous
differential equations \cite{ibra,Bluman}. The main characteristic of Lie
symmetry analysis is that it allows for the construction of one-parameter
point transformations which leave the given differential equation invariant.
From these point transformations, it is possible to derive similarity
transformations used to facilitate solutions. This procedure is known as the
reduction process, and in the case of partial differential equations, the
existence of a Lie symmetry is equivalent to the existence of a quantum operator.

For a review on the wide applications of symmetry analysis in classical
gravitational physics and cosmology, we refer the reader to \cite{nrev1}. In
the framework of the WDW equation, the mathematical approach for constructing
quantum operators from Lie symmetry analysis is analytically described in
\cite{abdb}. This method has been widely used for constructing the
wavefunction of the universe in a broad range of gravitational models
\cite{wd3,wd4,wd5,wd6,wd7,wd8,wd9,wd10}.

The application of the Lie symmetry analysis for the WDW equation
(\ref{rs.54}), i.e. the Yamabe equation, leads to the following quantum
operators%
\begin{align}
\Xi_{1}  &  :\left(  \partial_{\psi}-\frac{\beta_{1}}{\hbar}i\right)
\Phi=0,\\
\Xi_{2}  &  :\left(  -\frac{\left(  \kappa+1\right)  }{6}a\partial_{a}%
+\phi\partial_{\phi}-\frac{\beta_{2}}{\hbar}i\right)  \Phi=0,\\
\Xi_{3}  &  :\left(  \frac{\left(  \kappa+1\right)  }{6}\ln\phi a\partial
_{a}-\ln\phi~\phi\partial_{\phi}+\left(  \psi-\frac{2\left(  \kappa+1\right)
}{3}\ln a\right)  \partial_{\psi}-\left(  \frac{\left(  K-1\right)  }{8}%
\ln\phi+\frac{\beta_{3}}{\hbar}i\right)  \right)  \Phi=0.
\end{align}

With the use of the latter operators we are able to construct the following
solution for the wavefunction of the universe,%
\begin{equation}
\Phi\left(  a,\phi,\psi\right)  =a^{-\frac{3}{4}+\frac{2}{3}\frac{i}{\hbar
}\beta_{1}\left(  \kappa+1\right)  }\phi^{-\frac{\kappa+1}{8}+\frac{i}{\hbar
}\lambda}e^{-i\hbar\beta_{1}\psi}\left(  \Phi_{0}^{1}J_{\bar{\lambda}}\left(
\frac{2}{3\hbar}i\sqrt{3V_{0}}a^{3}\phi^{\frac{\kappa+1}{2}}\right)  +\Phi
_{0}^{2}Y_{\bar{\lambda}}\left(  \frac{2}{3\hbar}i\sqrt{3V_{0}}a^{3}%
\phi^{\frac{\kappa+1}{2}}\right)  \right)  , \label{rs.55}%
\end{equation}
in which $J_{\bar{\lambda}}$,~$Y_{\bar{\lambda}}$ are the Bessel functions of
the first and second kind respectively; and constants $\lambda,~\bar{\lambda}$
are defined as%
\begin{align}
\lambda &  =\frac{1}{9}\left(  \beta_{1}+9\beta_{2}+\beta_{1}\kappa\left(
2+\kappa\right)  \right)  ,\\
\bar{\lambda}  &  =\frac{i}{9}\sqrt{\frac{\beta_{1}}{\hbar}\left(
4\frac{\beta_{1}}{\hbar}\kappa\left(  \kappa+2\right)  +9i\left(
\kappa-1\right)  -72\frac{\beta_{2}}{\hbar}\right)  }.
\end{align}

Wavefunctions of the form in expression (\ref{rs.55}) have been explicitly
derived before in the literature \cite{wd9,wd10}. Thus, the same analysis for
the construction of a Hilbert space can be applied. We conclude the discussion
here and shift our focus to the classical limit.

Recall that for large values of the argument of the Bessel functions, the
asymptotic solution is
\begin{equation}
\Phi\left(  a,\phi,\psi\right)  \simeq a^{-\frac{3}{2}+\frac{2}{3}\frac
{i}{\hbar}\beta_{1}\left(  \kappa+1\right)  }\phi^{-\frac{\kappa+1}{4}%
+\frac{i}{\hbar}\lambda}e^{-\frac{i}{\hbar}\beta_{1}\psi}\left(  \bar{\Phi
}_{0}^{1}\cos\left(  K\right)  +\Phi_{0}^{2}\sin\left(  K\right)  \right)  ,
\end{equation}
with $K=\frac{2}{3\hbar}i\sqrt{3V_{0}}a^{3}\phi^{\frac{\kappa+1}{2}}%
-\frac{\bar{\lambda}\pi}{2}-\frac{\pi}{4}$; while in the limit $a^{3}%
\phi^{\frac{\kappa+1}{2}}\rightarrow0$, the asymptotic behaviour of the
wavefunction reads%
\begin{equation}
\Phi\left(  a,\phi,\psi\right)  \simeq a^{-\frac{3}{4}+\frac{2}{3}i\hbar
\beta_{1}\left(  \kappa+1\right)  +3\bar{\lambda}}\phi^{-\frac{\kappa+1}%
{8}+\frac{i}{\hbar}\lambda-\frac{\kappa+1}{2}\bar{\lambda}}e^{-i\frac
{\beta_{1}}{\hbar}\psi}.
\end{equation}

In the following lines, we focus on the analysis of the semiclassical limit of
quantum cosmology.

\section{Semiclassical limit}

\label{sec6}

In the Madelung representation \cite{md11} of quantum mechanics, that is, in
the hydrodynamic approach, we write the wavefunction in the for $\Phi\left(
q\right)  =\Omega\left(  q\right)  e^{\frac{i}{\hbar}S\left(  q\right)  }$,
where $\Omega$ is the amplitude of the wavefunction. By replacing in the WDW
equation (\ref{Int5}) and separating the real and imaginary parts we end with
the
\begin{equation}
\frac{1}{2}\mathcal{G}^{AB}\left(  \frac{\partial S}{\partial q^{A}}\right)
\left(  \frac{\partial S}{\partial q^{B}}\right)  +\mathcal{U}(\mathbf{q}%
)+\hbar^{2}\left(  \frac{n-2}{8(n-1)}\mathcal{R}-\frac{1}{2\Omega}%
\Delta\left(  \Omega\right)  \right)  =0. \label{ftb.54}%
\end{equation}

In the limit where $\hbar^{2}\rightarrow0$, that is, in the WKB approximation,
the Hamilton-Jacobi equation for the\ classical \ gravitational field
equations is recovered.

The new term $V_{Q}=-\frac{1}{2\Omega}\Delta\left(  \Omega\right)  $ is known
as the quantum potential in the de Broglie-Bohm representation of quantum
mechanics \cite{bm1,bm2} and it depends only on the amplitude of the
wavefunction. Furthermore, from the above expression it is clear how the
curvature of the minisuperspace contributes in the semiclassical limit.

The observables,\ $Q_{1},~Q_{2}$ and $Q_{3}$, lead to the conservation laws
for the classical system%
\begin{align}
I_{1}  &  =P_{\psi},\\
I_{2}  &  =-\frac{\left(  \kappa+1\right)  }{6}aP_{a}+\phi P_{\phi},\\
I_{3}  &  =\ln\phi\left(  \frac{\left(  \kappa+1\right)  }{6}aP_{a}-\phi
P_{\phi}\right)  +\left(  \psi-\frac{2\left(  \kappa+1\right)  }{3}\ln
a\right)  P_{\psi}-\frac{\left(  K-1\right)  }{8}\ln\phi.
\end{align}
in which $P_{a}=\frac{\partial S}{\partial a},~P_{\phi}=\frac{\partial
S}{\partial\phi}$ and $P_{\psi}=\frac{\partial S}{\partial\psi}$.

\subsection{Classical solution}

The Hamilton-Jacobi equation for the classical system reads%
\[
\frac{1}{6a\phi}\left(  \frac{\partial S}{\partial a}\right)  ^{2}+\frac
{4}{3a^{3}}\left(  \frac{\partial S}{\partial\phi}\right)  \left(
\frac{\partial S}{\partial\psi}\right)  +2V_{0}a^{3}\phi^{\kappa}=0\text{.}%
\]

By using the conservation laws we determine the closed-form solution for the
action $S\left(  a,\phi,\psi\right)  $, that is,
\begin{align}
S\left(  a,\phi,\psi\right)   &  =\frac{1}{9}\int\frac{9I_{2}+\sqrt{\left(
\kappa+1\right)  ^{2}\left(  I_{1}\left(  \left(  \kappa+1\right)  ^{2}%
I_{1}-18I_{2}\right)  -27V_{0}a^{6}\phi^{\kappa+1}\right)  }-\left(
\kappa+1\right)  ^{2}I_{1}}{\phi}d\phi\nonumber\\
&  +\frac{2}{3\left(  \kappa+1\right)  }\int\frac{\sqrt{\left(  \kappa
+1\right)  ^{2}\left(  I_{1}\left(  \left(  \kappa+1\right)  ^{2}I_{1}%
-18I_{2}\right)  -27V_{0}a^{6}\phi^{\kappa+1}\right)  }-\left(  \kappa
+1\right)  ^{2}I_{1}}{a}da\nonumber\\
&  +9U_{0}\left(  \kappa+1\right)  \int\int\frac{\phi^{\kappa}}{\sqrt{\left(
\kappa+1\right)  ^{2}\left(  I_{1}\left(  \left(  \kappa+1\right)  ^{2}%
I_{1}-18I_{2}\right)  -27V_{0}a^{6}\phi^{\kappa+1}\right)  }}d\phi
~da+I_{1}\psi. \label{ras.01}%
\end{align}

Thus
\begin{align}
P_{a}  &  =-\frac{2\left(  I_{1}\left(  \kappa+1\right)  ^{2}-\sqrt{\left(
\kappa+1\right)  ^{2}\left(  I_{1}\left(  \left(  \kappa+1\right)  ^{2}%
I_{1}-18I_{2}\right)  -27V_{0}a^{6}\phi^{\kappa+1}\right)  }\right)
}{3a\left(  \kappa+1\right)  },\\
P_{\phi}  &  =-\frac{I_{1}\left(  \kappa+1\right)  ^{2}-9I_{2}-\sqrt{\left(
\kappa+1\right)  ^{2}\left(  I_{1}\left(  \left(  \kappa+1\right)  ^{2}%
I_{1}-18I_{2}\right)  -27V_{0}a^{6}\phi^{\kappa+1}\right)  }}{9\phi},\\
P_{\psi}  &  =I_{1},
\end{align}
By replacing the momentum terms from (\ref{rs.51}) we end with a system of
three nonlinear first-order differential equations.

In the limit where $a^{6}\phi^{\kappa+1}\rightarrow0$, the field equations are
described by the following system (we have assumed the lapse function to be a
constant)%
\begin{align*}
\phi a^{2}\dot{a}  &  \simeq-\frac{A_{1}\left(  I_{1},I_{2},\kappa\right)
}{6},\\
\phi a^{3}\dot{\psi}  &  \simeq-\frac{2A_{2}\left(  I_{1},I_{2},\kappa\right)
}{27},\\
a^{3}\dot{\phi}  &  \simeq\frac{2}{3}I_{1}.
\end{align*}
where%
\begin{align*}
A_{1}\left(  I_{1},I_{2},\kappa\right)   &  =\frac{2}{3\left(  \kappa
+1\right)  }\left(  I_{1}\left(  \kappa+1\right)  ^{2}-\sqrt{\left(
\kappa+1\right)  ^{2}\left(  I_{1}\left(  \left(  \kappa+1\right)  ^{2}%
I_{1}-18I_{2}\right)  \right)  }\right)  ,\\
A_{2}\left(  I_{1},I_{2},\kappa\right)   &  =I_{1}\left(  \kappa+1\right)
^{2}-9I_{2}-\sqrt{\left(  \kappa+1\right)  ^{2}\left(  I_{1}\left(  \left(
\kappa+1\right)  ^{2}I_{1}-18I_{2}\right)  \right)  }.
\end{align*}

We write the equivalent system%
\begin{equation}
\int\frac{1}{\phi}d\phi\simeq4\frac{I_{1}}{A_{1}}\int\frac{1}{a}da,~\int
d\psi\simeq\frac{4A_{2}}{9A_{1}}\int\frac{1}{a}da,
\end{equation}
where the analytic solution is terms of the scale factor reads%
\begin{equation}
\phi\left(  a\right)  \simeq a^{\frac{4I_{1}}{A_{1}}}~,~\psi\left(  a\right)
\simeq\frac{4A_{2}}{9A_{1}}\ln a\text{.}%
\end{equation}

On the other hand, when the term $a^{6}\phi^{\kappa+1}$ dominates we end with
the following system%
\begin{align}
\dot{a}  &  \simeq-\frac{\sqrt{-\frac{3}{2}V_{0}}}{6\left(  \kappa+1\right)
}a\phi^{\frac{\kappa-1}{2}},\\
\dot{\psi}  &  \simeq-\frac{\sqrt{-2V_{0}}}{\sqrt{27}}\phi^{\frac{\kappa-1}%
{2}},\\
\dot{\phi}  &  \simeq\frac{2}{3}I_{1}a^{-3}.
\end{align}
The solution is real only when $V_{0}<0$. Thus, in this case in a similar way
as before we end with the reduced system%
\begin{align}
\int\phi^{\frac{\kappa-1}{2}}d\phi &  \simeq-\frac{4\sqrt{2}\left(
\kappa+1\right)  }{\sqrt{3\left\vert V_{0}\right\vert }}I_{1}\int a^{-4}da~,\\
\psi\left(  a\right)   &  \simeq\int\frac{4\left(  \kappa+1\right)  }{3}%
\phi^{\frac{\kappa-1}{2}}\frac{1}{a}da~.
\end{align}
which means that the asymptotic solution is%
\begin{equation}
\phi\left(  a\right)  \simeq\left(  -\frac{4\sqrt{2}\left(  \kappa+1\right)
I_{1}}{\sqrt{3\left\vert V_{0}\right\vert }}\right)  ^{\frac{2}{\kappa+1}%
}a^{-\frac{6}{\kappa+1}}~,~\psi\left(  a\right)  \simeq a^{3\frac{1-\kappa
}{1+\kappa}}.
\end{equation}

Recall that this solution is valid when $a^{3}\phi^{\frac{\kappa+1}{2}%
}\rightarrow0$, i.e. $\left(  -\frac{4\sqrt{2}\left(  \kappa+1\right)  I_{1}%
}{\sqrt{3\left\vert V_{0}\right\vert }}\right)  ^{\frac{2}{\kappa+1}%
}\rightarrow0$.

\subsubsection{Physical properties of the asymptotic solutions}

In the following lines we investigate the physical properties of the
asymptotic solutions derived before.

From the field equations of connection $\Gamma^{B},$ it follows that
\begin{equation}
H^{2}=\frac{1}{6\phi}\left(  V\left(  \phi\right)  -3\dot{\phi}\dot{\psi
}\right)
\end{equation}

Hence, for the asymptotic solution at the limit $a^{6}\phi^{\kappa
+1}\rightarrow0$, we derive%
\begin{equation}
H^{2}\simeq H_{0}^{1}\left(  I_{1},I_{2},\kappa\right)  a^{-6-8\frac{I_{1}%
}{I_{2}}}+H_{0}^{1}\left(  I_{1},I_{2},\kappa,V_{0}\right)  a^{4\frac{I_{1}%
}{I_{2}}\left(  1+\kappa\right)  -8\frac{I_{1}}{I_{2}}}.
\end{equation}
This corresponds to a cosmological solution where the effective fluid in the
framework of STEGR corresponds to two perfect fluids with constant equation parameters.

On the other hand, in the limit where the term $a^{6}\phi^{\kappa+1}$
dominates the asymptotic behavior for the Hubble function reads%
\begin{equation}
H^{2}\simeq\bar{H}_{0}^{1}\left(  I_{1},I_{2},\kappa,V_{0}\right)
a^{-6+\frac{12}{1+\kappa}}.
\end{equation}
The latter describes the perfect fluid solution with constant equation of
state parameter, which leads to a self-similar spacetime \cite{ndself}. The
equation of state parameter is defined as $w_{eff}=3-\frac{8}{1+\kappa}$, from
where it follows that acceleration is occurred when $-1<\kappa<\frac{5}{3}$.

\subsection{Quantum potential}

We determine the solution of the field equations in the semiclassical regime
where the quantum effects take place and affects the cosmological dynamics.

From the wavefunction (\ref{rs.55}) we calculate the quantum potential%
\begin{equation}
V_{Q}\left(  a,\phi,\psi\right)  =\frac{V_{Q}^{0}\left(  I_{1},I_{2}%
,\kappa\right)  }{\phi a^{3}},
\end{equation}
in which $V_{Q}^{0}\left(  I_{1},I_{2},\kappa\right)  $ is a constant related
to the parameters $I_{1}\,,~I_{2}$ and $\kappa$.

Consequently, the Hamilton-Jacobi equation (\ref{ftb.54}) becomes%
\[
\frac{1}{6a\phi}\left(  \frac{\partial\hat{S}}{\partial a}\right)  ^{2}%
+\frac{4}{3a^{3}}\left(  \frac{\partial\hat{S}}{\partial\phi}\right)  \left(
\frac{\partial\hat{S}}{\partial\psi}\right)  +2V_{0}a^{3}\phi^{\kappa}%
+\frac{2\bar{V}_{Q}^{0}\left(  I_{1},I_{2},\kappa\right)  }{a^{3}\phi}%
\hbar^{2}=0.
\]
The conservation laws presented before are also conservation laws and for the
latter Hamiltonian system with the quantum correction.

Thus, the action reads%
\begin{align}
\hat{S}\left(  a,\phi,\psi\right)   &  =\frac{1}{9}\int\frac{9I_{2}%
+\sqrt{\left(  \kappa+1\right)  ^{2}\left(  I_{1}\left(  \left(
\kappa+1\right)  ^{2}I_{1}-18I_{2}\right)  -27V_{0}a^{6}\phi^{\kappa+1}%
-27\bar{V}_{Q}^{0}\hbar^{2}\right)  }-\left(  \kappa+1\right)  ^{2}I_{1}}%
{\phi}d\phi\nonumber\\
&  +\frac{2}{3\left(  \kappa+1\right)  }\int\frac{\sqrt{\left(  \kappa
+1\right)  ^{2}\left(  I_{1}\left(  \left(  \kappa+1\right)  ^{2}I_{1}%
-18I_{2}\right)  -27V_{0}a^{6}\phi^{\kappa+1}-27\bar{V}_{Q}^{0}\hbar
^{2}\right)  }-\left(  \kappa+1\right)  ^{2}I_{1}}{a}da\nonumber\\
&  +9U_{0}\left(  \kappa+1\right)  \int\int\frac{\phi^{\kappa}}{\sqrt{\left(
\kappa+1\right)  ^{2}\left(  I_{1}\left(  \left(  \kappa+1\right)  ^{2}%
I_{1}-18I_{2}\right)  -27V_{0}a^{6}\phi^{\kappa+1}-27\bar{V}_{Q}^{0}\hbar
^{2}\right)  }}d\phi~da+I_{1}\psi.
\end{align}
We observe that the latter action is of the same function form of
(\ref{ras.01}) resulting the same gravitational field equations as before,
with a rescale on the constants $I_{1},~I_{2}$ and $\kappa$.

Consequently, the behaviour of the semiclassical solution is similar with that
derived before for the classical solution without the quantum correction term.

\section{Conclusions}

\label{sec7}

In this study, we reviewed the basic mathematical properties of nonmetricity
$f\left(  Q\right)  $-gravity. This gravitational model represents the
simplest generalization of STEGR by introducing nonlinear terms of the
nonmetricity scalar $Q$ into the gravitational Action Integral. $f\left(
Q\right)  $-gravity is a particular case of the more general
nonmetricity-scalar theory, where a scalar field is nonminimally coupled to
gravity. $f\left(  Q\right)  $-gravity exhibits Machian properties, even
though it is not purely Machian. Due to this characteristic, we discussed the
effects of conformal transformations on the gravitational Action Integral and
introduced the analogy of the Jordan and Einstein frames in nonmetricity theory.

Within the cosmological framework of FLRW geometry, the theory provides four
different sets of gravitational field equations. This arises because the
connection in STEGR theory is not uniquely defined. The choice of connection
leads to the introduction of geometrodynamical degrees of freedom in the field
equations, which affects the dynamics of the cosmological parameters. For the
coordinate system where the line element for FLRW is expressed in the usual
form of (\ref{rs.28}), there is a family of connections defined in the
so-called coincidence gauge. In this case, the connection depends on a gauge
function, which does not affect the cosmological dynamics. However, for the
remaining three families of connections, the geometrodynamical degrees of
freedom of the resulting gravitational equations can be attributed to two
scalar fields.~These two scalar field have a geometric origin related to the
dynamical degrees of freedom introduced by the nonlinear function $f$, and the
degrees of freedom introduced by the connection. ~

For the connection where the two scalar fields define a canonical kinetic
term, such that the gravitational field equations admit a minisuperspace
description, we employed the quantization process to derive the WDW equation
of quantum cosmology. We considered the power-law $f\left(  Q\right)
=f_{0}Q^{\mu}$ gravity, thus quantum observables were calculated using the
theory of Lie symmetries. In this consideration we focused in a multi-scalar
field cosmological model, either if the origin of the scalar field is geometric.

The quantum operators constructed by the Lie symmetry analsysis were used to
reduce the WDW equation and write a closed-form solution for the wavefunction.
We focused on the effects of quantum correction terms in the semiclassical
limit. Therefore, we calculated the action by solving the Hamilton-Jacobi
equation for the classical system, both with and without quantum correction
terms. From the derivation of the Action we were able to write the equivalent
reduced classical system.

We found that the quantum corrections do not affect the general evolution of
the classical solution. However, the quantum potential can rescale the
integration constants of the classical solutions. This is an important
observation because it states that the initial value problem can be overcome
by using the quantum corrections in the semi-classical limit.

Although $f\left(  Q\right)  $-cosmology is challenged by strong coupling and
the presence of ghosts, this geometric model provides a mathematical framework
for a better understanding of the effects of connection choice in
gravitational physics. Nonmetricity $f\left(  Q\right)  $-theory distinguishes
the connection from the metric, in contrast to GR, leading to new physics. For
example, there exists the gravitational model of dipole cosmology in a
Kantowski-Sachs background \cite{fqdip}. In this model, the tilted parameter
interacts with the dynamical degrees of the connection.

In future work, we plan to extend this analysis to study quantum corrections
in the semiclassical limit in the context of nonmetricity-scalar theory and
other extensions or modifications of STEGR.

\begin{acknowledgments}
AP thanks the support of VRIDT through Resoluci\'{o}n VRIDT No. 096/2022 and
Resoluci\'{o}n VRIDT No. 098/2022. Part of this work was supported by Proyecto
Fondecyt Regular 2024, Folio 1240514, Etapa 2024.
\end{acknowledgments}

\end{document}